\documentclass[twocolumn,showpacs,amsmath,amssymb,aps,prl]{revtex4}
 
\pdfoutput=1

\usepackage{graphicx}

\def\be{\begin{equation}}
\def\ee{\end{equation}}
\def\bea{\begin{eqnarray}}
\def\eea{\end{eqnarray}}

\def\s{\sigma}
\def\Tr{{\rm Tr}}
\def\l{\lambda}
\def\lm{{\l_{\rm m}}}

\begin{document}
\pagestyle{empty}

\title{Entanglement entropy of two disjoint blocks in critical Ising models}

\author{Vincenzo Alba$^1$, Luca Tagliacozzo$^2$, and Pasquale Calabrese$^3$}
\affiliation{$^1$ Scuola Normale Superiore and INFN, Pisa, Italy.
$^2$ 
School of Physical Sciences, the University of Queensland, QLD 4072, Australia.
$^3$Dipartimento di Fisica dell'Universit\`a di Pisa and INFN,  Pisa, Italy.}

\date{\today}

\begin{abstract}
We study the scaling of the  R\'enyi and entanglement entropy of two disjoint blocks  of critical Ising models,  
as function of  their sizes and separations.
We present analytic results based on conformal field theory that are quantitatively checked in
numerical simulations of both the quantum  spin chain  and the classical two dimensional Ising model.
Theoretical results match the ones obtained from numerical simulations only after taking properly into
account  the corrections induced by the finite length of the blocks to their leading scaling behavior.

\end{abstract}


\maketitle

Conformal field theory (CFT) is one of the most powerful and elegant tools to study quantum 
one-dimensional (1D) systems and classical two-dimensional (2D) ones. 
It provides a complete description of the low-energy (large-distance) physics of critical 
systems that can be classified only on the base of their symmetries \cite{cftbook}.
One spectacular recent success was the application of this framework to 
2D turbolence \cite{turb}. The predictions of CFT have been tested in experiments
for carbon nanotubes \cite{cn}, spin chains \cite{sc}, and cold atomic 
gases \cite{cg}, just to cite a few of the most recent ones.

CFT has been traditionally applied to the  computation of large
distance correlations of local observables.
Only recently it has been realized that CFT is also the ideal
tool to describe the global properties of a large subset of microscopical constituents 
(e.g. spins) and in 
particular their {\it entanglement}. This has generated an enormous interest in the study of 
the entanglement properties of many-body systems  \cite{rev} 
that is connecting several branches of physics such as quantum information, 
condensed matter, black hole physics. 
The quantum information insight about the origin of the achievements of the density matrix
renormalization group (DMRG) in 1D, and its failure in higher
dimensions \cite{Vidal},  can be cited as an example of  the outstanding
results generated by this cross-over between different branches of
physics.
The entanglement between two complementary regions A and  B of a
quantum system described by the state $ |\psi\rangle$ can be
measured through the entanglement entropy. This is
defined as  the  von-Neumann entropy of the reduced density matrix
$\rho_A=\Tr_B |\psi\rangle\langle\psi|$  obtained by 
tracing over the degrees of freedom in the region B. 
When $\psi$ is the ground state of an infinite 1D critical
system and A is a block of length $\ell$,  CFT  predicts the universal scaling \cite{Vidal,Holzhey,cc-04}
 \be
 S_A=\frac{c}3 \log \ell +c'_1\,,
 \label{SA}
 \ee
where $c$ is the central charge and $c'_1$ a non universal constant.
This formula is the most effective way to calculate the main signature of the CFT (the 
central charge), and it can be used to identify the universality class of new models, as for 
example done in the Fibonacci chain \cite{fibo}.

The reason of this simple scaling in CFT is easily understood \cite{cc-04}. 
In fact, through a replica trick, $S_A$ can be interpreted as $-\partial_n \Tr \rho_A^n|_{n=1}$. 
For integer $n$, $\Tr \rho_A^n$ is the partition function on an $n$-sheeted 
Riemann surface with two branch points at the border of the interval $A$ that can be mapped 
to the plane by a conformal transformation. 
By studying the transformation of the stress-energy tensor under
this conformal mapping, one has that $\Tr \rho_A^n$ is the two-point correlation
function of some {\it twist-operators} that have scaling dimension $\Delta_n=c/24 (n-1/n)$, 
i.e. $\Tr \rho_A^n=c_n \ell^{-c/6 (n-1/n)}$. 
By analytically continuing this to complex $n$ and by taking the derivative in 1, we 
get Eq. (\ref{SA}). 
This reasoning also applies to the case of $N$ intervals: $\Tr\rho_A^n$ is the partition function
of a $n$-sheeted Riemann surface with $2N$ branch points, i.e.
a $2N$-point function of the same twist-operators. A generally incorrect 
result was obtained by uniformizing this surface \cite{cc-04}. This is not allowed because of
the non-zero genus of the Riemann surface. 
This result was checked in several free-fermionic
theories \cite{ff} and only recently, the error has been pointed out \cite{cg-08,fps-09,cct-09}.
In the case of many intervals, $\Tr\rho_A^n$  turns out to be a function of the full operator 
content of the theory and not only of the central charge. 
For a free compactified boson or Luttinger liquid (LL) $\Tr\rho_A^n$ has been calculated 
for $n=2$ \cite{fps-09} and for general integer $n$ \cite{cct-09}. 
However, the functional dependence on $n$ is 
so complicated that the analytic continuation has not yet been achieved.
These predictions have been checked against the exact diagonalization
of the XXZ chain \cite{fps-09,cct-09}. 
Unfortunately, the numerical results are limited to relatively small system sizes
and only few general properties (like the dependence on the LL parameter) 
have been checked:  large oscillating corrections to the scaling (as
for one block \cite{ccn-09}), have made impossible a quantitative comparison for the 
scaling functions related to $\Tr \rho_A^n$. 
Concepts and calculation schemes used to get these results 
(such as higher genus Riemann surfaces, twist fields, orbifold theories) are mathematical tools that have been mainly used in string theory
and that only now find their place in condensed matter physics.

The entanglement of many intervals thus depends on the details of the CFT and should be 
calculated case by case \cite{footneg}. 
The simplest and most studied CFT is the critical Ising model that 
in the continuum is a free Majorana fermion and has central charge $c=1/2$. 
The corresponding 1D quantum spin chain is the Ising model in transverse field
described by the Hamiltonian
\be
H=-\sum_{j=1}^L [\s_j^x \s_{j+1}^x+h \s_j^z]\,,
\ee
where $\s_j^{x,z}$ are Pauli matrices acting on the spin at site $j$ and we use
periodic boundary conditions.
The model has a quantum critical point  at $h=1$.
The correspondence with a 
free fermion could erroneously lead to the conclusion that $S_A$ for the 
Ising chain is the incorrect result of Ref. \cite{cc-04}, valid for free 
fermion theories \cite{ff}. 
This  is not the case when the block $A$ involves more than one interval since the unitary
 transformation that maps the spin degrees of freedom to the fermionic ones is not anymore 
 contained inside $A$, as it is easily checked by direct calculation \cite{f-pr}. 
$S_A$ for two intervals has been 
calculated in the Ising chain \cite{ffip-08}, but for the fermion degrees of freedom and it
agrees with Ref. \cite{cc-04}.
The breaking of the equivalence of fermions and spins makes any lattice exact computation 
hard, and a representation of $\rho_A$ for two blocks is not yet known.
For this reason, we analyze the problem with numerical methods.
We use a tree tensor network (TTN) algorithm \cite{ttn} for the quantum 1D Ising model 
\cite{foot} 
and MonteCarlo simulations of the classical 2D one as in Ref.
\cite{cg-08}. 
Using the mapping to the torus partition function for $n=2$, 
we provide the CFT prediction for $\Tr\rho_A^2$. 
The generalization of this result to all integer $n$ requires a more detailed analysis
(as for the LL \cite{cct-09}, but more difficult because of the 
complexity of the target space \cite{Dixon,z-87}) that we are currently
studying and will be reported elsewhere \cite{c-prep}.

We consider the case of two disjoint intervals $A=[u_1,u_2]\cup [u_3,u_4]$. 
By global conformal invariance  $\Tr \rho_A^n$ can always be written as \cite{fps-09,cct-09}
\be
\Tr \rho_A^n
=c_n^2 \left(\frac{u_{31}u_{42}}{u_{21}u_{32}u_{43}u_{41}} 
\right)^{\frac{c}6(n-\frac1n)}  F_{n}(x)\,,
\label{Fn}
\ee 
where $u_{ij}=u_i-u_j$ and $x=u_{21}u_{43}/(u_{31} u_{42})$ is the four-point ratio.
$F_n(x)$ is the universal scaling function that depends on the theory, and $c_n$ the 
non-universal factor of the single block.
The normalization is $F_{n}(0)=1$. The incorrect result of Ref. \cite{cc-04} is $F_n(x)=1$
identically. 
For a chain of finite length $L$, one replaces $u_{ij}$ by the chord distance $L/\pi \sin(\pi u_{ij}/L)$.
$F_n(x)$ is symmetric for $x\to 1-x$ \cite{fps-09}. 

The TTN (as the better known DMRG) 
gives the full spectrum of the reduced density matrix. From this
$S_A$ and the moments of $\rho_A$ can be extracted and analyzed.
The scaling functions $F_n(x)$ (for the entropy $F_{VN}(x)=-F'_1(x)$) 
are obtained as ratios (difference) of $\Tr\rho_A^n$ ($S_A$) with the prefactor in Eq. (\ref{Fn}).
We consider two blocks of length $\ell$ at distance $r$.
The four-point ratio $x$ is obtained by substituting in its definition the chord distance:
\be
x=\left(\frac{\sin\pi\ell/L}{\sin\pi (\ell+r)/L}\right)^2\,.
\ee
In the $x$ variable, we would  expect that data with different  $\ell$, $r$ and $L$
 would collapse onto  a single curve thus revealing the scaling functions 
$F_n(x)$.

\begin{figure}[t]
\includegraphics[width=0.48\textwidth]{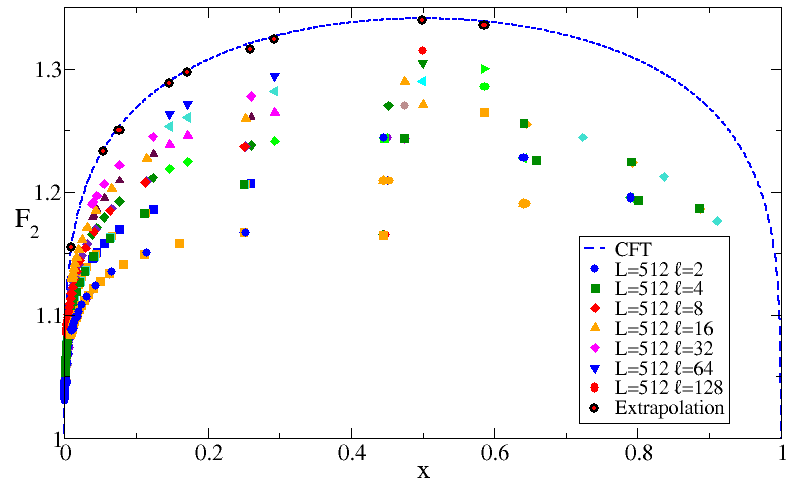}
\caption{TTN scaling function $F_2(x)$ vs the conformal ratio $x$ for different block sizes $\ell$.
The upper points are the extrapolation to 
$\ell\to\infty$ from Eq. (\ref{corr}). Data for $L\neq 512$ are not shown in the legend.
The full line is the CFT prediction (\ref{CFTF2}).}
\label{F2}
\end{figure}

\begin{figure}[t]
\includegraphics[width=0.48\textwidth]{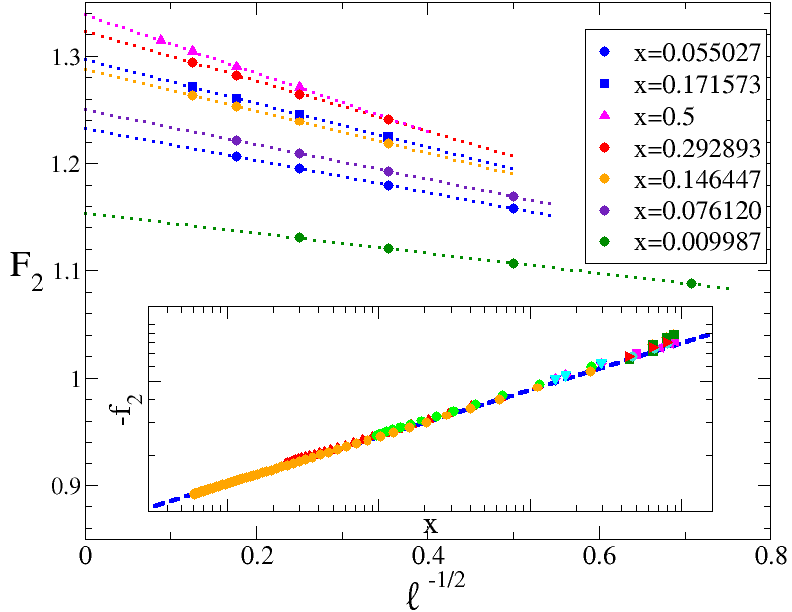}
\caption{Corrections to the scaling for $F_2^{\rm lat}(x)$  at fixed $x$ in Eq. \ref{corr}. 
Inset: universality of $f_2(x)$. The dashed line is  $\propto x^{1/4}$.
}
\label{corrfig}
\end{figure}

We start our analysis from the data for the function $F_2(x)$ reported in Fig. \ref{F2} for 
$\ell$ between $2$ and $128$ and $L$ from $64$ to $512$. 
The finite $\ell$ results do not display the symmetry
$x\to 1-x$ and the data present large corrections to their leading scaling behavior.
To extract the asymptotic behavior we perform 
a finite-size analysis. For any $x$, general RG arguments give the scaling
\be
F_2^{\rm lat}(x,\ell)= F_2^{CFT}(x)+\ell^{-\delta_c} f_2(x)+\dots\,,
\label{corr}
\ee
where $\delta_c$ is an unknown exponent, $f_2(x)$ is the scaling function of the first 
sub-leading correction, and the dots 
indicate further ones. The data are well described by $\delta_c=1/2$.
The evidence of this scaling for different $x$ is shown in Fig. \ref{corrfig}.
It is easy to extrapolate to $\ell\to \infty$ (the points where the straight lines 
cross the vertical axis) and the results are reported in 
Fig. \ref{F2}. 
The extrapolation restores the symmetry $x\to 1-x$. 
It is possible to calculate this quantity from CFT. In fact, the $2$-sheeted Riemann
surface has the topology of the torus, on which it can be mapped by a conformal 
transformation. 
The torus partition function for the Ising model is 
$2 Z^2_{\rm torus}=(\sum_{\nu=2}^4 |\theta_\nu(\tau)/\eta(\tau)|)^2$ \cite{cftbook}, 
where $\eta(\tau)$ is the Dedekin
function, $\theta_\nu(\tau)$ are the Jacobi elliptic functions and $\tau$ is the modular
parameter. In our case, $\tau$ is given by the solution of $x=[\theta_2(\tau)/\theta_3(\tau)]^4$
 \cite{Dixon}.  
 For this value of $\tau$, major simplifications occur (as for $\eta=1/2$ in the LL
 \cite{fps-09}) and the final result can be written in terms of 
only algebraic functions:
\begin{multline}
F_2(x) = \frac1{\sqrt{2}}\Bigg[
\left(\frac{(1 + \sqrt{x}) (1 + \sqrt{1 - x})}2\right)^{1/2} \\ 
+ x^{1/4} + ((1 - x) x)^{1/4} + (1 - x)^{1/4} \Bigg]^{1/2}\,.
          \label{CFTF2}
\end{multline}
This curve is reported in Fig. \ref{F2} and agrees with incredible precision with the
extrapolated data. For $x\ll1$ we have $F_2(x)=1+x^{1/4}/2+\dots$.
In the inset of Fig. \ref{corrfig}, we report the universal correction to the scaling function 
$f_2(x)$ obtained as $[F_2^{\rm lat}(x,\ell)- F_2^{CFT}(x)]\ell^{1/2}$ for 
different $\ell$ that collapse (without any adjustable parameter) on a single curve.
In the inset we show $f_2(x)\sim  x^{1/4}$.

\begin{figure}[t]
\includegraphics[width=0.48\textwidth]{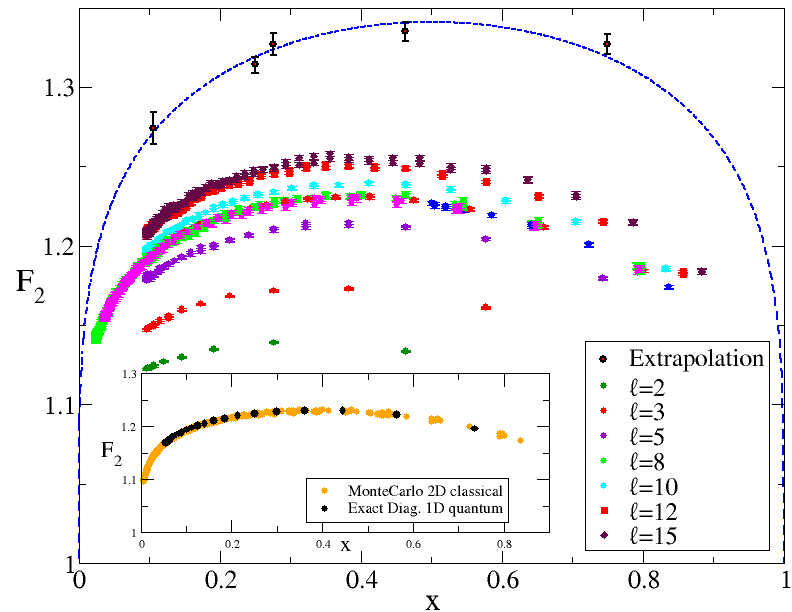}
\caption{MonteCarlo determination of the scaling function $F_2(x)$. 
The full line is the CFT prediction (\ref{CFTF2}).
Inset: Comparison between MonteCarlo data for the 2D classical Ising model and the 
exact diagonalization of the quantum chain. 
}
\label{MC}
\end{figure}

To check the universality,  we study the classical critical 2D Ising model, 
using the algorithm of Caraglio-Gliozzi to obtain the 
two-point function of twist-fields \cite{cg-08}. 
We use an asymmetrical geometry with the temporal direction $L_T$ equal to 10 times
the spatial one $L$ (between 24 and 324).
The results for $F_2(x)$ are reported in Fig. \ref{MC}
showing the same qualitative features as Fig. \ref{F2}. 
The extrapolations to $\ell\to\infty$ present large error bars, but in agreement with CFT. 
This also implies that a rescaling of all (large enough) length scales should give 
the same numbers in the two models (as in 2D  \cite{Gliozzi}).
The rescaling factor $a$ can be calculated from the single block entanglement obtaining
 $L_{2D}= a L_{1D}$, with $a\simeq 0.71$.
In the inset of Fig. \ref{MC}, the MonteCarlo data for the $L=8$ classical systems are compared with 
the $L=6 (\sim 8\times 0.71 )$ quantum chain showing a good agreement.

\begin{figure}[b]
\includegraphics[width=0.48\textwidth]{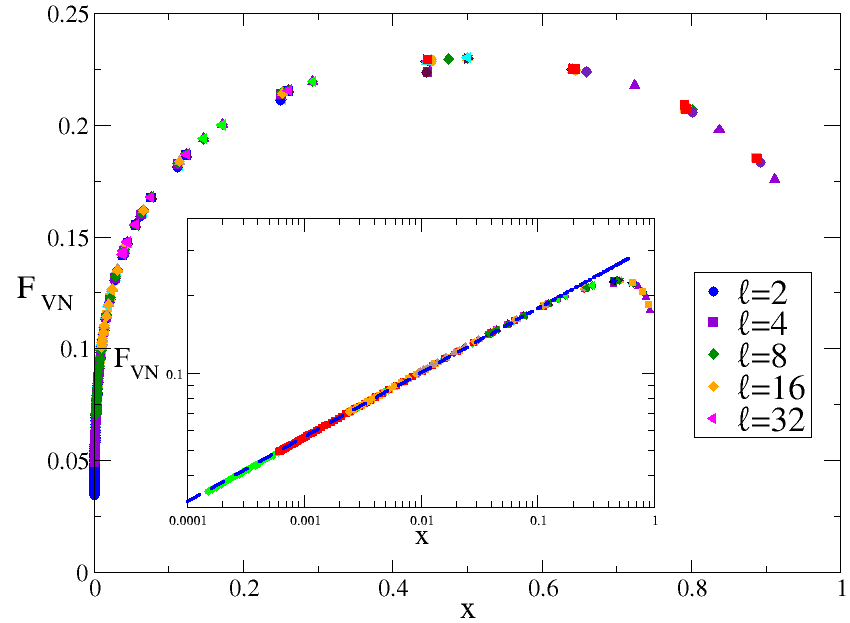}
\caption{TTN data for the scaling function $F_{VN}(x)$.
Corrections to the scaling are negligible and all data collapse. 
Inset: Same data in log-log scale showing the power-law behavior for small $x$ with 
the predicted exponent $1/4$ and the prefactor $\pi$. 
}
\label{VN}
\end{figure}

In Fig. \ref{VN} we report the TTN scaling function for $F_{VN}(x)$. 
Unfortunately the CFT value is unknown because we are not yet able to make 
the analytic continuation (as for the LL).  
One important feature is evident from the plot: the corrections to the scaling 
are negligible and all data collapse in a single symmetric scaling curve. 
In the inset of the figure we report the data in log-log scale to emphasize the power-law 
behavior for small $x$. 
In the LL, $F_n(x)$ for small $x$
displays a power-law with an $n$-independent exponent  \cite{cct-09}. 
This reasoning generalizes to the Ising model \cite{c-prep}
and from the result for $F_2(x)$ we read that the exponent is $1/4$, as confirmed by 
the plot. We also found that the prefactor is $\pi$. 
Moreover, for various $n$, we computed the function $F_n(x)$ for the $n$-th moment  of 
$\rho_A$ also showing large finite $\ell$ corrections. 
The analysis of these data will be reported elsewhere \cite{c-prep}.

\begin{figure}[t]
\includegraphics[width=0.48\textwidth]{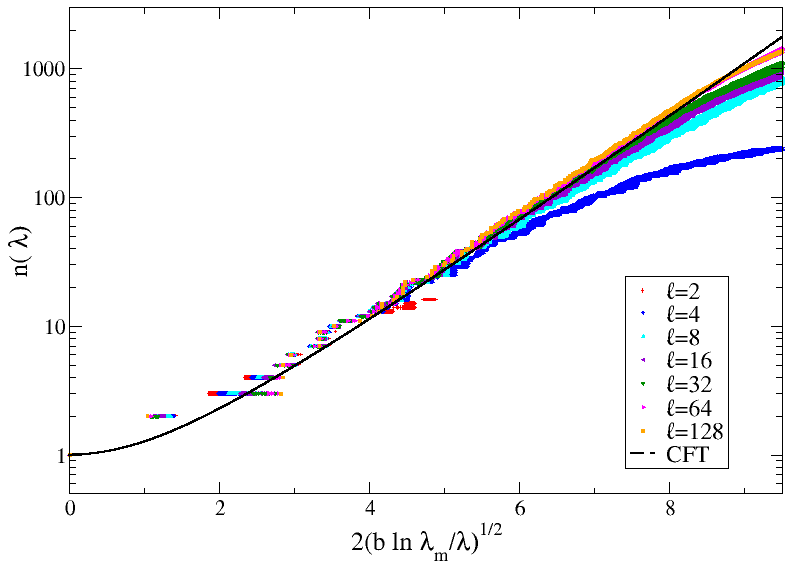}
\caption{TTN spectrum of the reduced density matrix.  
In the scaling variable of the horizontal axis all data collapse on the CFT prediction (\ref{nlam}).}
\label{spectrum}
\end{figure}

Finally, we consider the full spectrum of $\rho_A$. 
If  the moments of $\rho_A$ behave
like $\Tr\rho_A^n\simeq L_{\rm eff}^{-c/6(n-1/n)}$ with a prefactor roughly independent on $n$, 
then the spectrum displays the super-universal (i.e. independent on any details of the theory) form
\cite{cl-08}
\be
n(\l)=\int_\l^{\l_{\rm m}} d\l P(\l) =
I_0(2\sqrt{b \ln (\lm/\l)})\,,
\label{nlam}
\ee
where $n(\l)$ is the mean number of eigenvalues larger than $\l$, ${\l_{\rm m}}$ the 
maximum eigenvalue, $b=-\ln \l_{\rm m}$, and $I_0(y)$ a Bessel function. 
This implies that if $n(\l)$ is plotted against $y=2\sqrt{b \ln (\lm/\l)}$ all data of any system 
should collapse on the same curve.  
In Fig. \ref{spectrum} we plot $n(\l)$ against $y$ and all TTN data at 
different $L,\ell,r$ (for a total of more than $10^5$ points) collapse on the 
curve predicted by CFT. Finite size effects are present for small $\ell$. 
Such good agreement is due to the 
fact that $c_n^2 F_n(x)$ slightly depends on $n$, varying by few per cents in the 
range $[2,\infty]$. This spectrum is  fundamental 
to describe the scaling of numerical algorithms \cite{t-08}.

To summarize, we reported a full analytic and numerical analysis of the entanglement 
of two disjoint intervals in the Ising universality class. This represents  the first  
numerical check of the CFT predictions (also derived in this letter) 
for quantities that are more complicated than the entanglement of the single block. 
It would be interesting to understand how these results change in systems with boundaries
(that already for the single interval present intriguing features \cite{cc-04,lsca-06}) 
and in the presence of quenched disorder, to understand if the apparent ``restoration'' of 
conformal invariance for one interval \cite{rm-04} is somehow preserved in the case of many.

{\it Acknowledgments}.
We are grateful to M. Fagotti, F. Gliozzi,  E. Tonni, and G. Vidal for discussions.


\begin{thebibliography}{99}
\bibitem{foot}
The TTN  naturally encodes the RG ideas. Tensors are organized in layers corresponding 
to different length scales  \cite{ttn}. This makes it  particularly suited for finite size studies. 
Indeed,   passing from  a smaller system to a bigger one only requires the optimization of  a single 
tensor on the top of the network corresponding to the new length scale.
\end{thebibliography}
\end{document}